\documentclass[graybox,vecphys]{svmult}

\usepackage{mathptmx}       
\usepackage{helvet}         
\usepackage{courier}        
\usepackage{type1cm}        
\usepackage{makeidx}         
\usepackage{graphicx}        
\usepackage{multicol}        
\usepackage[bottom]{footmisc}

\makeindex             

\begin{document}

\title*{An F-statistic based multi-detector veto for detector artifacts in continuous-wave gravitational wave data}
\author{D. Keitel, R. Prix, M.A. Papa, M. Siddiqi}
\institute{Albert-Einstein-Institut, Hannover and Golm, Germany. \email{david.keitel@aei.mpg.de}}
%
%
\maketitle

\abstract*{Continuous gravitational waves (CW) are expected from spinning neutron stars with non-axisymmetric deformations. A network of interferometric detectors (LIGO, Virgo and GEO600) is looking for these signals. They are predicted to be very weak and retrievable only by integration over long observation times. One of the standard methods of CW data analysis is the multi-detector F-statistic. In a typical search, the F-statistic is computed over a range in frequency, spin-down and sky position, and the candidates with highest F values are kept for further analysis. However, this detection statistic is susceptible to a class of noise artifacts, strong monochromatic lines in a single detector. By assuming an extended noise model - standard Gaussian noise plus single-detector lines - we can use a Bayesian odds ratio to derive a generalized detection statistic, the line veto (LV-) statistic. In the absence of lines, it behaves similarly to the F-statistic, but it is more robust against line artifacts. In the past, ad-hoc post-processing vetoes have been implemented in searches to remove these artifacts. Here we provide a systematic framework to develop and benchmark this class of vetoes. We present our results from testing this LV-statistic on simulated data.}

\abstract{Continuous gravitational waves (CW) are expected from spinning neutron stars with non-axisymmetric deformations. A network of interferometric detectors (LIGO, Virgo and GEO600) is looking for these signals. They are predicted to be very weak and retrievable only by integration over long observation times. One of the standard methods of CW data analysis is the multi-detector $\mathcal{F}$-statistic. In a typical search, the $\mathcal{F}$-statistic is computed over a range in frequency, spin-down and sky position, and the candidates with highest $\mathcal{F}$ values are kept for further analysis. However, this detection statistic is susceptible to a class of noise artifacts, strong monochromatic lines in a single detector. By assuming an extended noise model - standard Gaussian noise plus single-detector lines - we can use a Bayesian odds ratio to derive a generalized detection statistic, the line veto (LV-) statistic. In the absence of lines, it behaves similarly to the $\mathcal{F}$-statistic, but it is more robust against line artifacts. In the past, ad-hoc post-processing vetoes have been implemented in searches to remove these artifacts. Here we provide a systematic framework to develop and benchmark this class of vetoes. We present our results from testing this LV-statistic on simulated data.}

\vspace{0.5cm}

 In a search for gravitational waves, we are conducting hypothesis tests: at a certain point in parameter space (frequency, spin-down and sky position), is there a signal or not? Assuming Gaussian detector noise only, we have two hypotheses, $\mathcal{H}_{\mathrm{G}} : \vec{x}(t) = \vec{n}(t)$ and $\mathcal{H}_{\mathrm{S}} : \vec{x}(t) = \vec{n}(t) + \vec{h}(t,\mathcal{A})$, where $\mathcal{A}$ are additional signal parameters, like polarization angles. In the Bayesian approach, we compute the odds ratio of the two hypotheses, and we marginalize over the unknown parameters $\mathcal{A}$:
 \begin{equation}
  O_{\mathrm{SG}}(\mathbf{x}) \equiv \frac{P\left(\mathcal{H}_{\mathrm{S}}|\vec{x}\right)}{P\left(\mathcal{H}_{\mathrm{G}}|\vec{x}\right)} \propto \int \frac{P\left(\vec{x}|\mathcal{H}_{\mathrm{S}},\mathcal{A}\right)}{P\left(\vec{x}|\mathcal{H}_{\mathrm{G}}\right)} P\left(\mathcal{A}|\mathcal{H}_{\mathrm{S}}\right) \mathrm{d}\mathcal{A}
 \end{equation}
 The marginalization can be done analytically (for specific priors on $\mathcal{A}$, see \cite{pgm11, PrixKrish09}). We obtain $O_{\mathrm{SG}}(\mathbf{x}) \propto \mathrm{e}^{\mathcal{F}(\vec{x})}$, with the standard multi-detector $\mathcal{F}$-statistic \cite{cs05,jks98}.

 The problem with this approach is that quasi-monochromatic, stationary detector artifacts (''lines'') look more like $\mathcal{H}_{\mathrm{S}}$ than $\mathcal{H}_{\mathrm{G}}$ and will result in large values for $O_{\mathrm{SG}}$. So we add an alternative noise hypothesis $\mathcal{H}_{\mathrm{L}}$ that fits lines in single detectors better than the multi-detector coherent $\mathcal{H}_{\mathrm{S}}$, namely $\mathcal{H}_{\mathrm{L}}^X : \vec{x}^X(t) = \vec{n}^X(t) + \vec{h}^X(t,\mathcal{A})$ for a signal in only one detector $X$, but pure noise $\mathcal{H}_{\mathrm{G}}$ in the others. Again using the $\mathcal{F}$-statistic priors and analytically maximizing over $\mathcal{A}$, we can (e.g. for two detectors $X=1,2$) replace the standard $\mathcal{F}$-statistic by a new detection statistic with an extended noise hypothesis:
 \begin{equation}
   O_{\mathrm{SN}}(\mathbf{x}) \equiv \frac{P\left(\mathcal{H}_{\mathrm{S}}|\vec{x}\right)}{P\left(\mathcal{H}_{\mathrm{L}}|\vec{x}\right) + P\left(\mathcal{H}_{\mathrm{G}}|\vec{x}\right)}
   \propto \frac{\mathrm{e}^{\mathcal{F}(\vec{x})}}{\rho_{\mathrm{max}}^4/70 + l^1\,\mathrm{e}^{\mathcal{F}^1(\vec{x}^1)} + l^2\,\mathrm{e}^{\mathcal{F}^2(\vec{x}^2)}} \label{eq:lv_odds_sn}
 \end{equation}
 The new detection statistic downweights candidates which have higher single-detector than multi-detector $\mathcal{F}$-statistics, thereby penalizing lines. The $l^X$ are the prior line probabilities, while the parameter $\rho_{\mathrm{max}}$ from a signal strength prior allows us to tune the detection statistic, determining how much discrepancy between detectors is attributed to Gaussian noise and how soon vetoing sets in. Further work on simulated data is necessary to choose this prior optimally.

 In preliminary studies with simulated data, we found the new detection statistic to be much more effective than the standard semi-coherent $\mathcal{F}$-statistic, as seen in the figure below. Especially at low false-alarm rates, which are desirable for GW searches, the new statistic allows for more detections. See \cite{pkpls} for more details.

 \begin{center}
  \includegraphics[width=0.4\textwidth,angle=270]{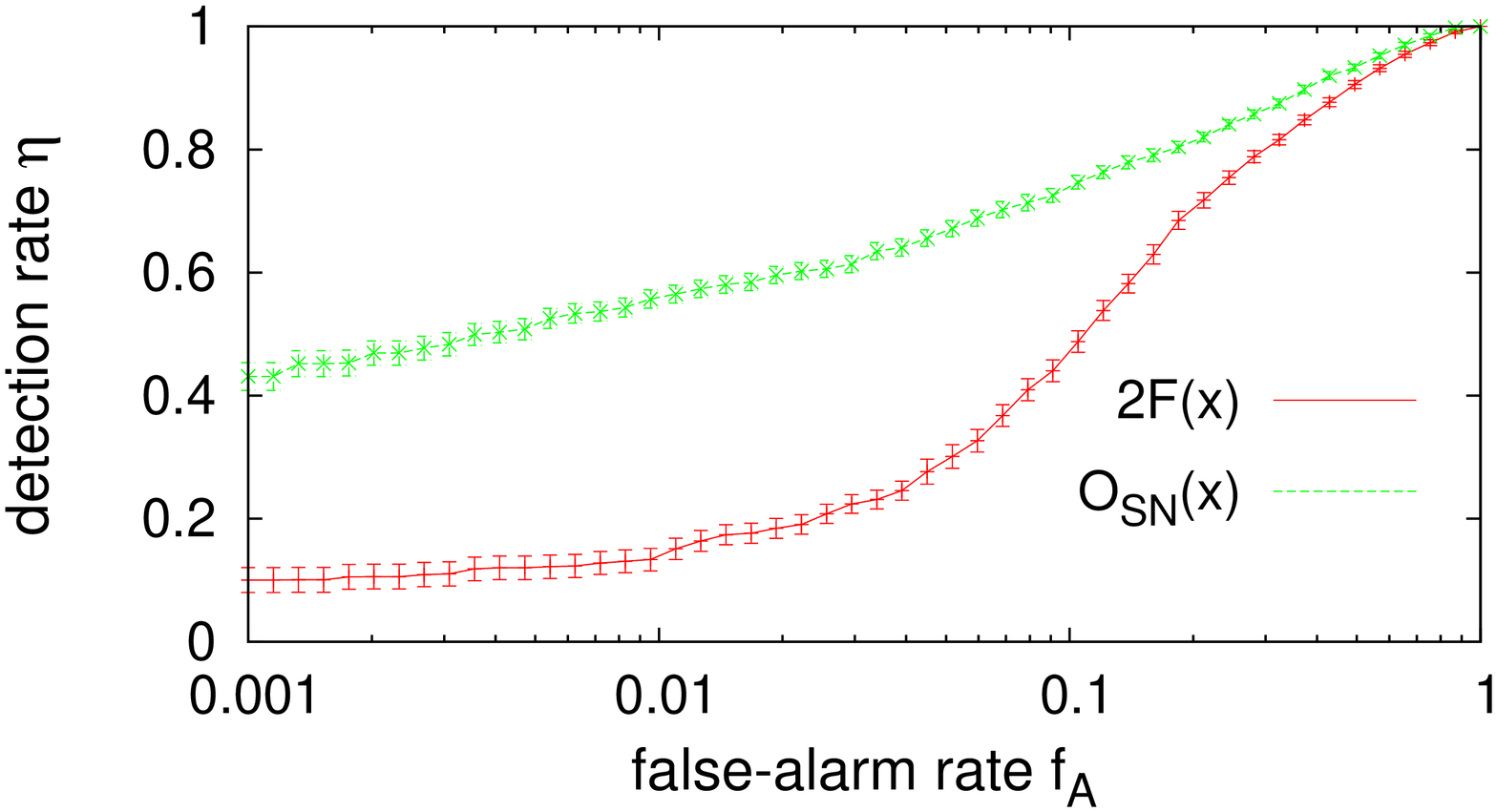} \\
 \end{center}

\end{document}